\documentclass[draft]{agujournal2019}
\usepackage{url} 
\usepackage{lineno}
\usepackage[inline]{trackchanges} 
\usepackage{soul}

\draftfalse

\journalname{Radio Science}

\begin{document}

\title{Analytic approximations of scattering effects on beam chromaticity in 21-cm global experiments}


\authors{Alan E. E. Rogers\affil{1}, John P. Barrett\affil{1}, Judd D. Bowman\affil{2}, Rigel Cappallo\affil{1}, Colin J. Lonsdale\affil{1}, Nivedita Mahesh\affil{2}, Raul A. Monsalve\affil{3,2,5}, Steven G. Murray\affil{2}, Peter H. Sims\affil{4}}

\affiliation{1}{Haystack Observatory, Massachusetts Institute of Technology, Westford, MA 01886, USA}
\affiliation{2}{School of Earth and Space Exploration, Arizona State University, Tempe, AZ 85287, USA}
\affiliation{3}{Space Sciences Laboratory, University of California, Berkeley, CA 94720, USA}
\affiliation{4}{McGill Space Institute and Department of Physics, McGill University, Montreal, QC, Canada}
\affiliation{5}{Facultad de Ingenier\'ia, Universidad Cat\'olica de la Sant\'isima Concepci\'on, Alonso de Ribera 2850, Concepci\'on, Chile}

\correspondingauthor{Alan E. E. Rogers}{arogers@haystack.mit.edu}

\begin{keypoints}
\item In order to accurately measure the spectrum of the radio sky the antenna beam needs to be smooth without frequency structure 
\item The beam of the antenna on its ground plane is influenced by the scatter from nearby objects which produces ripples in the spectrum
\item Analytic expressions provide an estimate of the scatter in an environment too complex for accurate electromagnetic modeling of the beam
\end{keypoints}


\begin{abstract}

Scattering from objects near an antenna produce correlated signals from strong compact radio sources in a manner similar to
those used by the “Sea Interferometer” to measure the radio source positions using the fine frequency structure in the total
power spectrum of a single antenna. These fringes or ripples due to correlated signal interference are present at a low level in the spectrum of any single antenna
and are a major source of systematics in systems used to measure the global redshifted 21-cm signal from the early universe. In the Sea Interferometer a single antenna on a
cliff above the sea is used to add the signal from the direct path to the signal from the path reflected from the sea thereby forming an
interferometer. This was used for mapping radio sources with a single antenna by Bolton and Slee in the 1950s. In this paper we derive analytic expressions
to determine the level of these ripples and compare these results in a few simple cases with electromagnetic modeling software
to verify that the analytic calculations are sufficient to obtain the magnitude of the scattering effects on the measurements of the global 21-cm signal.
These analytic calculations are needed to evaluate the magnitude of the effects in cases that are either too complex or take too much time to be modeled using software.

\end{abstract}

\section{Introduction} 

The spectrum of the radio sky in the 50 to 200 MHz frequency band is relatively smooth because there are no strong spectral lines. When the sky is observed with a small antenna with a large smooth beam on an infinite ground plane the
observed spectrum is the average  of many continuum radio sources and will be smooth over frequency if the receiver and antenna reflection coefficient are well calibrated.
However, in practice, sky noise from compact sources will be scattered
from objects like trees, bushes, rocks, uneven ground and other antennas
surrounding the ground plane. Raised areas of the ground plane
also act as scatterers.
In this case the signals from the radio sources that are not excluded by a zero response in the antenna beam in the
direction of the scattering object can also enter the antenna from a separate path as shown in Figure 1. In this case
the correlations between the signals in the direct and scattered paths form ripples in the total power spectrum as in the Sea Interferometer (Bolton \& Slee, 1953).
These spectral ripples can cause problems for inference of the 21-cm signal, which relies on the a priori
assumption of foregrounds remaining spectrally smooth. We derive simple analytic expressions that approximate this scattering effect, which should be useful in modeling the expected
systematics of global experiments but electromagnetic modeling software is needed for the inclusion of significant objects in the beam calculation.

\begin{figure}
\noindent\includegraphics[width=\textwidth]{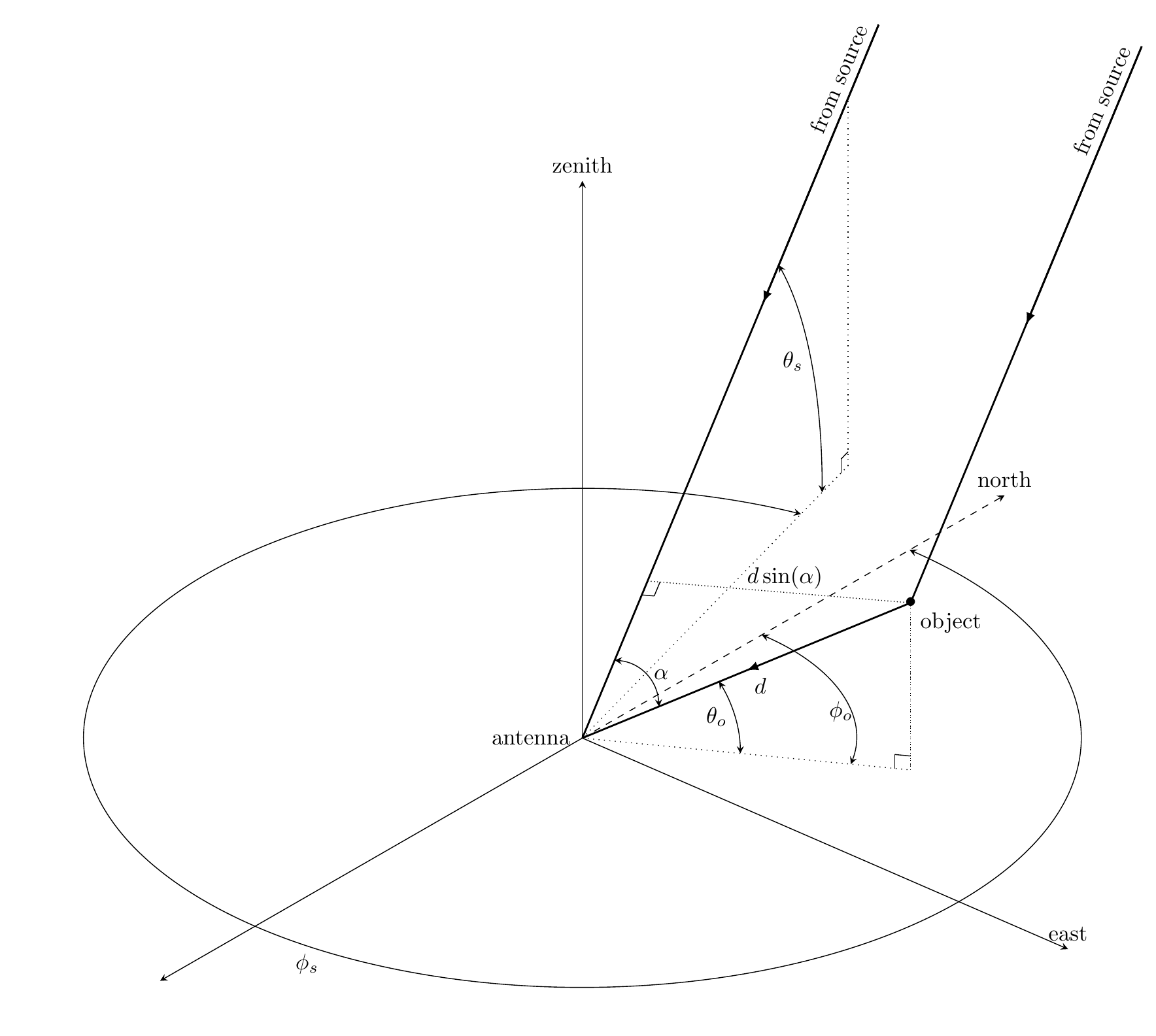}
\caption{Diagram of the geometry of the antenna, source, and object, showing the direct path from the radio source and the indirect path via the scattering object to the antenna. The antenna is at the origin and the object is a distance $d$ away at (elevation, azimuth) = $(\theta_{o}, \phi_{o})$. The radio source is at infinity at ($\theta_{s}$, $\phi_{s})$. The angle, $\alpha$, between the direction to the source and the direction to the object is given by $\alpha = \arccos(\mathbf{\hat{r}_s} \cdot \mathbf{\hat{r}_o} )$ where $\mathbf{\hat{r}_s}$ and $\mathbf{\hat{r}_o}$ are vectors from the antenna to the source and object.}
\end{figure}

\section{Algorithms} 
In this section we derive an analytic approximation for the ripple fraction induced by a single source.
The power, $P_{source}$, received directly from the source is given by

\begin{equation}
 P_{source} = A_{source}F
\end{equation}
\\
where $A_{source}$ is the effective aperture area of
the antenna on its ground plane in the direction of
the radio source with flux $F$.

 The power, $P_{object}$, scattered from an object with radar cross section, $\sigma$, is

\begin{equation}
 P_{object} = F\sigma
\end{equation}

While the radar cross section of a scattering object is usually approximated by its physical cross section the scattering cross section
is more accurately estimated using the Rayleigh scattering analysis. For example an object the size of the EDGES (Bowman et al., 2018)
electronics hut at the Murchison Radio Observatory (MRO), which is about $5m^2$ in physical cross section, has a radar cross section (RCS) (Knott et al. 2004) that is about $4$ times larger than the
physical cross section at 75 MHz because the hut circumference is close to a wavelength which is the peak of the Rayleigh scattering region.
For more complex objects of several wavelengths in size, like trees and bushes, the RCS is closer to the physical cross section and power is scattered more isotropically.

If the power is scattered isotropically the power, $P_{scat}$, received by the antenna is

\begin{equation}
 P_{scat} = A_{scat} \left ( \frac{F\sigma}{4\pi d^2} \right )
\end{equation}
\
where $d$ is the distance of the scatterer from the antenna and
$A_{scat}$ is the effective aperture (area) of the antenna in the direction of the scattering object.
On the assumption that the direct and scattered signals are perfectly correlated the combined complex voltage, $v$, is given by

\begin{equation}
  v = P_{source}^{ \frac{1}{2} } + e^{i\omega\tau}P_{scat}^{ \frac{1}{2} }
\end{equation}
\\
and the total received power $P_{total} = vv^*$. When $P_{scat}$ is small relative to $P_{source}$, then to first order $P_{total}$ is given by

\begin{equation}
 P_{total} = P_{source}\left\{ 1 + 2cos(\omega\tau) \left[ \frac{A_{scat}}{A_{source}} \frac{\sigma}{4\pi d^2} \right ]^{\frac{1}{2}} \right\}
\end{equation}
\\
where $\omega = 2\pi{f}$, $f$ is frequency and $\tau$ is the delay of the scattered signal relative to the direct signal. The geometry of the source, antenna, and scattering object is shown in Figure 1, which shows that $\tau = \frac{d}{c}(1-\cos\alpha)$, where $c$ is the speed of light, and $\alpha$ is the angle subtended by the arc from source to object as seen from the antenna. In terms of the (elevation, azimuth) of the source, $(\theta_s, \phi_s)$, and object, $(\theta_o, \phi_o)$, the delay $\tau$ can be expressed as

\begin{equation}
 \tau = \frac{d}{c} \bigg [ 1-\cos\Delta\cos\delta + \sin\theta_s\sin\theta_o (\cos\delta-1) \bigg ]
\end{equation}
\\
 where $\Delta = \theta_s - \theta_o$ and $\delta = \phi_s - \phi_o$. When the elevation angle, $\theta_o$, is small, then
$\tau \approx \frac{d}{c} (1-\cos\Delta\cos\delta)$ and reaches a maximum of $\frac{2d}{c}$ when the antenna is in the direction of the source as seen from the object.

The fractional ripple $r$ = $(P_{total}-P_{source})/P_{source}$ is given by
\begin{equation}
r = \pi^{- \frac{1}{2} }\cos(\omega\tau) \left( \frac{G_{scat}}{ G_{source} }\right)^{\frac{1}{2}} \left ( \frac{\sigma^{\frac{1}{2}}}{d} \right) 
\end{equation}
\\
where $G_{scat}$/$G_{source}$ is the ratio of the antenna gain in the direction of the scattering object
and the source which is equivalent to and has been substituted for the ratio of effective areas $A_{scat}/A_{source}$.

\section{Examples} \label{sec:examples}

As an example the flux of Cas A at 100 MHz is $1.4\times10^4$ Jy which results in an antenna temperature of 23 K
for $G_{source}$ = 8 dB typical at a high elevation. Then the substitution of $G_{scat}$ = -23 dB for the antenna gain in the direction of a scattering
object, which is a typical gain at less than 10 degrees elevation, with radar cross section of  $\sigma =10m^2$ at
$d$ = 75m results in a peak to peak ripple of 31 mK using Equation 7.
It is noted that the ripple magnitude decreases in proportion to the inverse of the distance but at large distances the source
may be resolved by the projected interferometric baseline, which has length ${dsin(\alpha)}$, thereby reducing the correlation. In addition the ripple period in frequency will be shortened so that
it may be appreciably smoothed by the spectral resolution bandwidth. It should also be noted that scattering by multiple objects will generate a more complicated structure in frequency. The ripple frequency period which is the inverse of $\tau$ given in Equation 6 can become very long for a radio source which is
low in elevation at the azimuth of the scattering object. In this case the scattering effects are more spread out over frequency and Galactic Hour Angle (GHA).

The example shown in Figure 2 is the result of the scattering of the Galactic center region by the electronics hut
which is about 50m from the EDGES (Bowman et al. 2018) lowband-1 antenna. This observation consists of data from day 250 in 2016 to day 95 in 2017, averaged over 10 minute blocks of GHA from 03:00 hours to 04:10. These data have been corrected using an antenna beam model (Mahesh et al. 2021) using the FEKO Method of Moments (MoM) software (https://altairhyperworks.com/product/FEKO) without the effects of the hut, and is compared to simulated data (solid line curves) produced using a FEKO beam model that includes the hut. In each case the Haslam sky map at 408 MHz (Haslam et al. 1982) scaled by a spectral index of -2.5 from 55 to 95 MHz was used for the convolution of the sky (Mozdzen et al. 2019) with the beam. The dashed line curves are obtained by adding the product of the
ripple fraction from Equation 7 by the value of each pixel of the Haslam map in the convolution with the FEKO beam model without the hut. Alternatively and equivalently every gain value of the FEKO beam without the hut can be multiplied by one plus the ripple fraction to create a beam that when convolved with the sky map yields the same dashed line curves. 
While the model using the ripple fraction does not agree precisely with the data, this is because the analytic approximations do not account for the full details, especially
the phase, of the scattering. However, they do have roughly the same root mean square (rms) magnitude as the solid line curves, and are therefore useful for estimating the order of magnitude of
the ripple systematic in complicated geometries that would be too difficult to model with electromagnetic software.

The fine frequency structure of the residuals in Figure 2 are the result of the fine frequency structure in the beam which is introduced by the correlations which change with the delay of the scattered signals from the hut in the same manner as the “fringes” of the Sea Interferometer.  The effect of a different spectral index from different point sources in the sky map is relatively small because the spectral index is a spectrally smooth function which is taken out by the 5-physical terms. The relatively large effect on the observed spectrum which results from scattering from objects in the environment of the antenna is the result of the fine frequency structure that is added to the beam. The ripple fraction in Equation 7 provides a means of approximation of the fine frequency structure added to the beam in the presence of scattering objects with minimal additional computation but ability to accurately model and remove the effects of the fine structure introduced by the scatter depends on a sky map with an accurate frequency dependence of the point sources and an accurate beam which includes the objects which produce significant scatter.

\begin{figure}
\noindent\includegraphics[width=\textwidth]{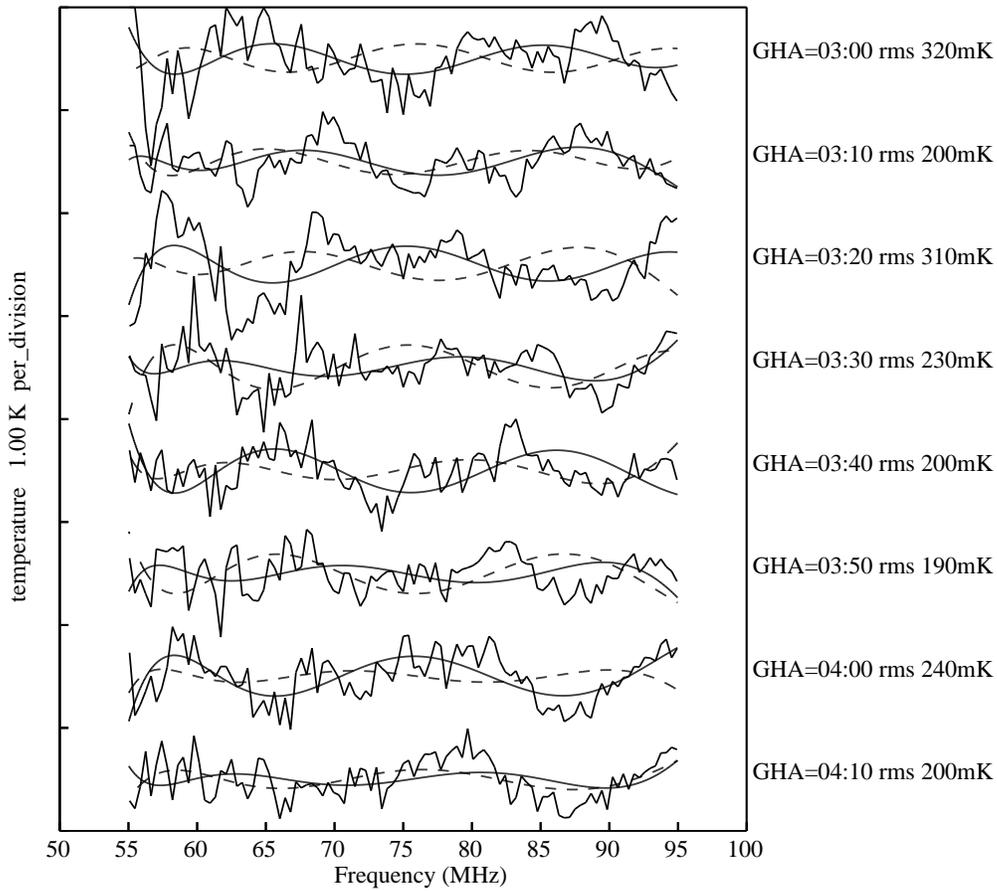}
\caption{Comparison of the residuals (after a model with 5-physical terms (Bowman et al. 2018) for the sky noise spectrum is removed) of beam corrected data and simulation. The simulation (in solid line curves) was produced using the lowband antenna beam modeled by FEKO assuming an infinite perfectly electrically conducting (PEC) ground plane and the hut 50 meters west of the antenna which is pointed north. The lowband-1 data (solid jagged curves with GHA time and rms listed on the right side of each plot) exhibits frequency structure consistent with scatter from the hut. The dashed lines are simulated using the ripple fraction from Equation 7 - see text.}
\end{figure}

\subsection{Table of scattering ripple amplitudes}

\begin{table}
\caption{Scattering examples of a cube with face area of $5m^2$ at different distances $d$ from the antenna on different ground planes at various latitudes.
$rms1$ and $rms2$ are the average of the rms values for each of the 24 1-hour blocks of GHA using the FEKO simulation and analytic approximation respectively. $rms3$ is the rms for
the scattering averaged over all 24 hours of GHA. The FEKO simulations used dielectric $3.5$ and conductivity $0.02$ S/m for soil and $80$ and $0.06$ S/m for a lake.}
 \centering
 \begin{tabular}{c c c c c c}
 \hline
        antenna      & latitude   &  d    & $rms1$ & $rms2$ &  $rms3$    \\
        ground plane & (deg)      & (m)   &  (mK)  &   (mK) &   (mK)     \\
 \hline
 MRO 30x30m          &  -27       &  50   &   84   &    78  &    16      \\
 lake vert. pol      &  -27       &  50   &  322   &   293  &    40      \\
 lake vert. pol      &  -27       & 100   &   47   &    64  &     9      \\
 horiz. 30x15m       &   68       &  30   &  121   &   116  &    22      \\
 horiz. 30x15m       &   42       &  30   &  112   &    98  &    16      \\
 \hline
 \end{tabular}
 \end{table}

In addition to comparing the analytic approximations with real data and electromagnetic simulations in Figure 2, FEKO
simulations are compared with the analytic approximations using the horizontally polarized EDGES dipole on the 30x30m
ground plane in (Mahesh et al. 2018) and a vertically polarized monopole antenna similar to that used
by SARAS (Singh et al. 2022) consisting of an inverted cone of radius 1m and height of 1m over a circular plate of 1m radius.

Table 1 shows the level of scattering for a single cubic scatterer at at distance $d$ from the antenna, representing a hut, with a physical cross section of $5m^2$
and a radar cross section of about $\sigma\approx20m^2$ when the scattering effects in the Rayleigh regime are taken into account using the actual physical model of the cube in the FEKO model.
The simulations using FEKO to obtain the average rms of the ripples for 24 1-hour blocks over all GHA from 55 to 95 MHz are given in $rms1$. The rms obtained by adding the
product of the ripple fraction from Equation 7 by the value of each pixel of the Haslam map above the horizon, which has been scaled by a spectral index of $-2.5$, in a convolution with
the FEKO beam model without the hut is used for the approximate estimate of the rms given in $rms2$. The rms residual for the effects of scatter averaged over
all 24 hours from the FEKO estimate is listed in $rms3$ and shows that averaging over 24 hours significantly reduces the net effects of the scattering.

The results are shown for different antennas at different sites. The third case is for the hut at 100m from
a vertically polarized antenna. While Equation 7 shows that the rms should only drop by a factor of two for a distance of 100
meters, the result in Table 1 shows a drop of more than a factor of 4 for both the FEKO modeling and the estimate, owing to the
increased angular resolution of the longer projected baseline and a more rapid change with GHA.
While different ground planes and the change of sky coverage with latitude, have some effect, in general the rms residuals from scattering are proportional
to the square root of the radar cross section and inversely
proportional to the distance of the scattering objects from the antenna. The effect of the soil on the antenna beam is accounted for in the FEKO model
and has only a small effect on the antenna gain factors in Equation 7 so that the effect on the fractional ripple estimate is not significant. It is also found with
FEKO model simulations that for a given antenna the ripple amplitude produced by a scattering
object at a fixed location relative to the antenna has only small dependence on the ground plane size and soil.
The scattering effects for vertical polarization are much larger than for horizontal polarization
because the relative gain of a vertically polarized antenna near the horizon is much higher than for a horizontally polarized antenna.
To limit the scattering effects when a vertically polarized antenna is used, a much larger ground plane and surrounding flat area are needed.
Averaging over only 12 hours of GHA also reduces the rms but by a smaller amount. An analysis of the same FEKO simulation used for the first entry of Table 1 gave average
values of 44 and 21 mK for 12 1-hour blocks centered at 0 and 12 hours GHA respectively, compared with 16 mK when averaging over 24 hours from the value of $rms3$
in the first entry of Table 1.

 \begin{table}
 \caption{The first case is a repeat of the last case in Table 1. The second case is a cube on the east
and the last case is the FEKO and analytic approximation of both  east and west cubes as an
example of multiple scatterers}
 \centering
 \begin{tabular}{c c c c c c}
 \hline
        antenna      & latitude & $d$   & $rms1$ & $rms2$  & $rms3$  \\
        ground plane & (deg)    & (m)   &  (mK)  &  (mK)   &  (mK)   \\
 \hline
 horiz. 30x15m       &  42      & 30W   &  112   &   98    &   16    \\
 horiz. 30x15m       &  42      & 30E   &   86   &   94    &   16    \\
 horiz. 30x15m       &  42      & 30W+E &  147   &  148    &   23    \\
 \hline
 \end{tabular}
 \end{table}

\subsection{Multiple scatterers}

The analytic expression for the ripple fraction in Equation 7 can be summed over multiple scattering objects as long as
the terms $\tau$ and $G_{source}$ which are dependent coordinates of each pixel in the sky map are recomputed for each pixel.
This allows the magnitude of the scattering effects of a complex environment surrounding the antenna and its ground plane to be easily
computed in cases for which running FEKO or other electromagnetic software is impractical. This summation should be valid for the case that the scattered signals
are from objects that are wavelengths apart so that the signals from each object are likely to be uncorrelated with each other. 
The last case in Table 1, which has a cube on east side of the antenna is repeated
for a cube on the west side and then for the sum of the ripple fraction of both cubes and the results are shown in Table 2. In order to assess a site for global 21-cm observations to avoid
being limited by scattering the ripple fraction should be summed over all objects out to 100 meters from the antenna.

\section{Summary and Conclusions}

We derive simple analytic expressions to assess the level of systematics in the spectra from a single antenna to show that observations of the
global 21-cm signal require a ground plane or flat area which is large enough to avoid objects which produce fine structure in the spectra via scattering.
These simulations and expressions show that a vertical polarized antenna is more sensitive to scattering objects. As a result the
ground plane or a flat area may have to extend out to 100 meters from the antenna, depending on the radar cross-section of scattering objects and low angle gain of the antenna,
to avoid being limited by scattering.


\acknowledgments

This work was supported by the NSF through research awards for the Experiment to Detect the Global EoR Signature
 (AST-0905990, AST-1207761, AST-1609450 and AST-1909307). We thank the Murchison Radio Observatory for their support.
We acknowledge the Wajarri Yamatji people as the traditional owners of the Observatory site. We thank
CSIRO for providing site infrastructure and support.

\section*{Data Availability Statement}

The data used in Figure 2 for the comparison against the analytic approximation of the scatter is publically available at https://loco.lab.asu.edu/edges/edges-data-release/

\clearpage











\end{document}